\documentclass[letter,twocolumn]{jpsj3}
\usepackage{amsfonts}
\usepackage{ulem}

\usepackage{graphicx}
\usepackage{amsmath}
\usepackage{braket}
\usepackage{bm}
\usepackage{color}
\usepackage{txfonts}

\begin{document}

\title{Topological Modes Protected by Chiral and Two-Fold Rotational Symmetry in a Spring-Mass Model with a Lieb Lattice Structure}
\author{Hiromasa Wakao$^1$, Tsuneya Yoshida$^2$, Tomonari Mizoguchi$^2$, and Yasuhiro Hatsugai$^2$}
\inst{%
$^1$Graduate School of Pure and Applied Sciences, University of Tsukuba, Tsukuba, Ibaraki 305-8571, Japan\\
$^2$Department of Physics, University of Tsukuba, Tsukuba, Ibaraki 305-8571, Japan
}%

\date{\today}

\abst{
We propose how to realize the topological modes, which correspond to topological zero modes for a quantum system, protected by chiral and rotation symmetry for a mechanical system.
Specifically, we show the emergence of topological modes protected by chiral and two-fold rotational symmetry by a spring-mass system with a Lieb lattice structure and dents on the floor. 
Moreover, comparing the results of a tight-binding model, we have found the additional topological modes for our spring-mass model due to the extra degrees of freedoms.
Our approach to realize the topological modes can be applied to other cases with rotation symmetry, e.g., a system of a honeycomb lattice with three-fold rotational symmetry.
}

\maketitle

{\bf Introduction:}~
In this decade, topological phases have been extensively analyzed as new quantum states which host boundary modes protected by topological properties in the bulk~\cite{Hatsugai1993,Ryu2002,Kane2005,Hatsugai2013,RevModPhys.82.3045,RevModPhys.83.1057,ando2013topological,PhysRevB.78.195125,PhysRevLett.106.106802,slager2013space,PhysRevB.81.134509,Ryu_2010,PhysRevB.84.153101,PhysRevB.86.115131,PhysRevLett.110.240404,PhysRevB.88.075142,RevModPhys.88.035005,PhysRevB.90.115207,PhysRevB.97.205135, Candido2018}. 
Remarkably, it turned out that these topological phenomena can be observed even beyond the quantum systems~\cite{kane2014topological,Wang_2015,kariyado2015manipulation,NatlAcadSciU.S.A.113.E4767,PhysRevB.94.125125,bertoldi2017flexible,takahashi2017edge,lee2018topolectrical,imhof2018topolectrical,PhysRevB.99.024102,RevModPhys.91.015006,Ota:19,PhysRevB.100.054109,arXiv:1912.12022,PhysRevB.101.094107} (e.g., photonic crystals~\cite{PhysRevB.94.125125,RevModPhys.91.015006,Ota:19}, mechanical systems~\cite{kane2014topological,Wang_2015,kariyado2015manipulation,NatlAcadSciU.S.A.113.E4767,bertoldi2017flexible,takahashi2017edge,PhysRevB.99.024102,PhysRevB.100.054109,PhysRevB.101.094107} etc.). 
These topological phenomena beyond the quantum systems originate from the fact that these classical systems are described by an eigenvalue equation.
One of the advantages of these classical systems is high controllability of their parameters.
So far thanks to their high controllability, a variety of topological phases has been proposed for spring-mass systems (SMMs) which are periodic arrangements of the mass points and springs~\cite{Wang_2015,kariyado2015manipulation, PhysRevB.99.024102,takahashi2017edge,PhysRevB.100.054109,PhysRevB.101.094107}.
For instance, Chern insulators~\cite{Wang_2015,kariyado2015manipulation} and higher-order topological phases~\cite{PhysRevB.101.094107} have been realized for SMMs.

Along with the above development, the notion of the topological protection has also been extended to gapless quantum systems. 
A representative example is a Weyl semi-metal showing the gapless modes in the bulk.
These gapless modes are protected by a finite value of the Chern number which is the topological invariant for two-dimensional systems in class A (no symmetry)~\cite{Weng_WeylTaAs_PRX15,Su_WeylFermion_Sci349}. 
Similar topological gapless modes can also be found for other local symmetry (e.g. time-reversal symmetry, particle-hole symmetry, and chiral symmetry)~\cite{Bzdusek_Nodal_PRB96}.
Furthermore, the analysis of topological gapless states is further developed by taking into account the spatial symmetry. 
In particular, the coexistence of chiral symmetry and spatial symmetry gives topologically stable zero energy modes protected by these symmetry~\cite{PhysRevB.90.115207}.

To observe the robustness of these topological modes, chiral symmetry with a high accuracy is needed. However, unfortunately, chiral symmetry for quantum systems such as metals and insulators can be broken due to the long range hoppings or the spin-orbit coupling, etc.

Under this background, in this paper, we theoretically propose how to realize the above topological zero modes by taking advantage of the high controllability of the classical systems.
Specifically, we realize the topological zero modes for the SMM with a Lieb lattice structure.
We note that the simple preparation of the SMM does not preserve chiral symmetry; by tuning the on-site potential arising from the dents on the floor, we have a SMM preserving chiral symmetry.
In addition, by comparing the results of a tight-binding model (TBM), we also find that the extra degrees of freedoms of the SMM increase the number of topological zero modes, which is a unique phenomenon of SMMs.
Our approach for the realization can be applied to a general two-dimensional tight-binding model (TBM).


{\bf The Topological Zero Modes Protected by Chiral and Spatial Symmetry:}~Firstly, we briefly review the general arguments of the topological zero modes protected by chiral and spatial symmetry for quantum systems~\cite{PhysRevB.90.115207}.
It is well-known that the Schr\"{o}dinger equation is reduced to the eigenvalue problem.
We set a matrix $H$ for the eigenvalue equation.
Additionally, we consider a chiral operator $\Upsilon$ and a spatial operator $R$.
Chiral symmetry and spatial symmetry of $H$ are written as
\begin{subequations}
 \begin{equation}
  \label{eq:chiral symmetry}
  \{H,\Upsilon\}=0,
 \end{equation}
 \begin{equation}
  \label{eq:spatial symmetry}
  [H,R]=0.
 \end{equation}
\end{subequations}
From (\ref{eq:chiral symmetry}), chiral symmetry makes the pair of positive and negative eigenvalues of $H$.

Let us assume that chirality of lattices is not changed by the operators $R$,
\begin{equation}
 \label{eq:asuumeption}
 [\Upsilon,R]=0.
\end{equation}
From (\ref{eq:spatial symmetry}) and (\ref{eq:asuumeption}), one can see that the matrices $H$, $\Upsilon$ and $R$ are block-diagonalized by the eigenvectors of $R$,
\begin{subequations}
 \begin{equation}
\label{eq:block H}
  H=
\begin{pmatrix}
 H_{R_1} & & \\
 & H_{R_2} & \\
 &  & \ddots
\end{pmatrix},
 \end{equation}
 \begin{equation}
\label{eq:block U}
  \Upsilon=
\begin{pmatrix}
 \Upsilon_{R_1} & & \\
 & \Upsilon_{R_2} & \\
 &  & \ddots
\end{pmatrix},
 \end{equation}
 \begin{equation}
\label{eq:block U}
  R=
\begin{pmatrix}
 R_1 & & \\
 & R_2 & \\
 &  & \ddots
\end{pmatrix},
 \end{equation}
 \begin{equation}
\label{eq:block R}
  R_{i}= r_{i} \hat{1}_{n_{i}} \quad \mathrm{for} \ i=1,2,\cdots,
 \end{equation}
\end{subequations}
where $R_i$, $r_{i}$ ($r_i \neq r_j$ for $i\neq j$ ), and $n_{i}$ denote an eigenspace of $R$, an eigenvalue, and the number of degeneracy of $r_{i}$, respectively.
Here, the indices of $H$ and $\Upsilon$ denote the eigenspace of $R$.
Additionally, $\hat{1}_{n}$ is the $n \times n$ identity matrix.
We note that chiral symmetry and spatial symmetry are preserved at each eigenspaces, $\{H_{R_i}, \Upsilon_{R_i}\}=0$ and $[H_{R_i}, R_i]=0$.

The topological index is defined as
\begin{equation}
 \nu_{R_i}= \mathrm{Tr}\Upsilon_{R_i},
\end{equation}
for each eigenspace~\cite{PhysRevB.90.115207,Sutherland1986}.
This index denotes the difference between the number of $u^{+}_{R_i}$ and $u^{-}_{R_i}$ where $Hu^{\pm}_{R_i}=0$ and $\Upsilon_{R_i}u^{\pm}_{R_i}= \pm u^{\pm}_{R_i}$.
Therefore, this index guarantees the $|\nu_{R_i}|$ topological zero modes protected by chiral symmetry in the eigenspace $R_{i}$ at least.
For the total system, the topological index is written as
\begin{equation}
 \nu = \sum_{R_{i}} |\nu_{R_i}|.
\end{equation}
This index guarantees that at least there exist $\nu$ the topological zero modes protected by the operators $R$ and $\Upsilon$ in the total system.
In the rest of this paper, by the topological zero modes, we denote topological zero modes protected by chiral symmetry and rotation symmetry.

\begin{figure}[htbp]
 \centering
 \includegraphics[width=0.8\linewidth]{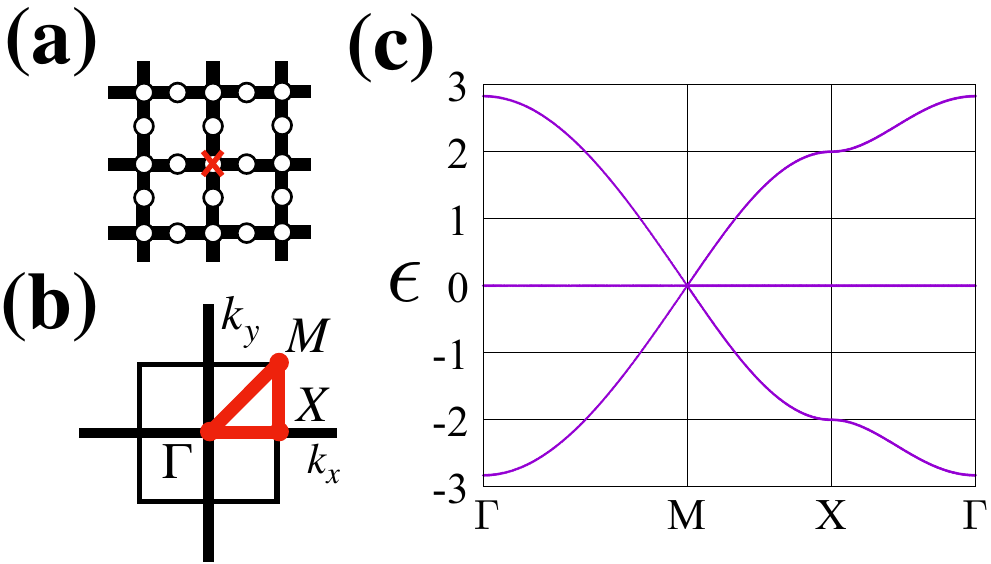}
 \caption{(Color online)
(a) A TBM of the Lieb lattice structure.
\text{\sffamily X} dentoes the center of the two-fold rotation.
(b) The first Brillouin zone of the TBM.
(c) The band structure for $t=1$ along the line shown in the panel (b).
}
 \label{fig:TBM}
\end{figure}

{\bf TBM with a Lieb Lattice Structure:}~
Based on the above generic argument, we analyze the nearest-neighbor TBM with a Lieb lattice structure under the periodic boundary condition [see Fig.~\ref{fig:TBM} (a)].
Our analysis for the TBM elucidates that the three-fold degeneracy at $M$ point is protected by chiral symmetry and two-fold rotational symmetry.
For a later use, we set unit vectors as $\vec{a}_{1}=(1,0)^T$ and $\vec{a}_{2}=(0,1)^T$.
High symmetry points $\Gamma$, $M$, and $X$ in the first Brillouin zone are located at $(k_x , k_y) =(0,0)$, $(\pi,\pi)$, and $(\pi,0)$, respectively [see Fig.~\ref{fig:TBM} (b)].

For the TBM, the problem is reduced to the eigenvalue equation, $h(\vec{k}) \vec{\psi}_{\vec{k}} = \epsilon \vec{\psi}_{\vec{k}}$ where $h(\vec{k})$, $\vec{\psi}_{\vec{k}}$, and $\epsilon$ denote the bulk hamiltonian, an eigenstate function, and an eigenenergy, respectively.
Here, $p=1,2,3$ represents the sublattice.
$\psi_{\vec{k};p}$ denotes the element of the eigenstate function $\psi_{\vec{k}}$.

Under the basis $(\psi_{\vec{k},1},\psi_{\vec{k},2},\psi_{\vec{k},3})$, the explicit form of the bulk Hamiltonian is written by
\begin{equation}
 \label{eq:bulk Hamiltonian}
  h(\vec{k})=
  \begin{pmatrix}
   0 & -t-te^{- i k_x} & -t - te^{-i k_y} \\
   -t-te^{i k_x} & 0 & 0 \\
   -t-te^{i k_y} & 0 & 0
  \end{pmatrix},
\end{equation}
where $t$ denotes the nearest-neighbor hopping parameter taking a real value.
By solving the eigenvalue equation of $h(\vec{k})$, we obtain the dispertion relation.
Figure~\ref{fig:TBM} (c) shows the band structure for $t=1$ in the first Brillouin zone.
Here, the horizontal axis denotes the high-symmetry lines in the Brillouin zone connecting $\Gamma$, $M$, and $X$ points [see Fig.~\ref{fig:TBM} (b)].

Let us focus on chiral symmetry and two-fold rotational symmetry of this system.
We introduce a chiral operator $\Upsilon_{T}$ and a two-fold rotational operator $R_{T}$ around the point denoted by \text{\sffamily X} in Fig.~\ref{fig:TBM} (a),
\begin{subequations}
\label{eq:operator}
 \begin{equation}
\label{eq:Chiral operator T}
 \Upsilon_{T}= \mathrm{diag}(1,-1,-1),
\end{equation}
\begin{equation}
\label{eq:Spatial operator T}
 R_{T}(\vec{k})= \mathrm{diag}(1,e^{-ik_x},e^{-ik_y}).
\end{equation}
\end{subequations}
Two-fold rotational symmetry is written as $R_{T}(\vec{k})h(\vec{k})R^{-1}_{T}(\vec{k})=h(-\vec{k})$.
From Eq.~(\ref{eq:bulk Hamiltonian}) and Eq.~(\ref{eq:operator}), 
one can confirm two equations $\{h(\vec{k}_{0}),\Upsilon_{T}\}=0$ and $[h(\vec{k}_{0}),R_{T}(\vec{k}_{0})]=0$ with $\vec{k}_{0}$ denoting momentum at high symmetry points $\Gamma$, $M$ and $X$.
This indicates the preservation of chiral
and two-fold rotational symmetry
at the high symmetry points $\Gamma$, $M$ and, $X$.

Here, we discuss topologically protected band touching at high-symmetry points.
From the Eq.~(\ref{eq:Chiral operator T}) and Eq.~(\ref{eq:Spatial operator T}), one can see that the equation $ [\Upsilon_{T},R_{T}(\vec{k})]=0$ holds.
This denotes that chirality of the lattices is not changed by $R_{T}$.
Therefore, $\Upsilon_{T}$ can be block-diagonalized by eigenvectors of $R_{T}$.
Thus, the number of the topological zero modes protected by two-fold rotational symmetry at the high-symmetry points is
\begin{equation}
 \nu =
  \begin{cases}
   1 & \mathrm{at} \quad \Gamma \\
   3 & \mathrm{at} \quad M \\
   1 & \mathrm{at} \quad X
  \end{cases}.
  \label{topological index for TBM}
\end{equation}
The above results can be obtained from $\Upsilon_{R_{1}=1}$ and $\Upsilon_{R_{2}=-1}$.
For example, at $M$ point, we have $\Upsilon_{R_{1}=1}=1$ and $\Upsilon_{R_{2}=-1}=\mathrm{diag}(-1,-1)$, which result in $\nu_{R_{1}=1}=1$ and $\nu_{R_{2}=-1}=-2$.  
Form Eq.~(\ref{topological index for TBM}), we can deduce that topological zero modes can be observed as one flat band and one Diarc cone at $M$ point, and as one flat band at $\Gamma$ and $X$ points.
In fact, this can be confirmed from the band structure [see Fig.~\ref{fig:TBM} (c)].
These topological zero modes are stable as long as this system preserves chiral and two-fold rotational symmetry since operators defined in Eq.~(\ref{eq:operator}) are not changed.
Here, we would like to stress that, to our konwledge, topological origin at $M$ point has not been discussed in terms of chiral symmetry and spatial symmetry.

In the above, we have discussed topological zero modes protected by chiral symmetry and two-fold symmetry for the TBM.
We note however, chiral symmetry with a high accuracy is needed to observe the robustness of topological zero modes protected by chiral symmetry; chiral symmetry for metals and insulators can be broken due to the long range hoppings or the spin-orbit coupling, etc.



\begin{figure}[htbp]
\centering
\includegraphics[width=\linewidth]{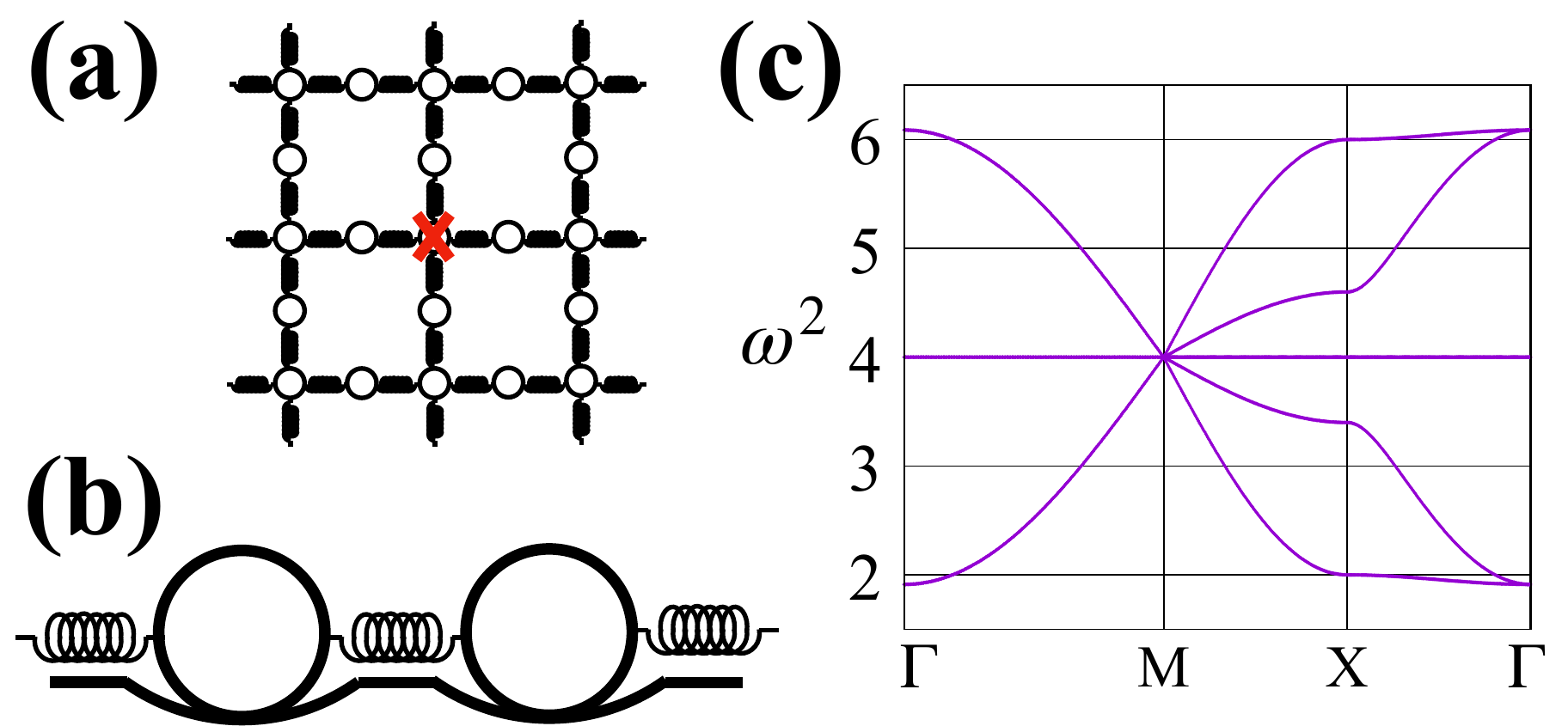}
\caption{(Color online)
(a) The SMM of the Lieb lattice structure.
\text{\sffamily X} denotes the center of the two-fold rotation.
(b) The mass points oscillating along the surface of the dent due to the gravity.
(c) The band structure along the line shown in the panel Fig.~\ref{fig:TBM} (b).
}
\label{fig:SMM}
\end{figure}

{\bf SMM with a Lieb Lattice Structure:}~
In contrast to the TBM, a SMM preserves chiral symmetry without any approximation since the coupling appears between the bases connected by springs.
Therefore, we realize the topological zero modes for a SMM with a Lieb lattice structure [see Fig.~\ref{fig:SMM} (a)] to overcome the difficulty in realizing the preservation of chiral symmetry for the TBM.

Let us see the motion of the mass points in the SMM with a Lieb lattice structure under the periodic boundary condition~\cite{kariyado2015manipulation}.
For the simplicity, we set the mass of mass points as unit, and unit vectors as $\vec{a}_{1}=(1,0)^T$ and $\vec{a}_{2}=(0,1)^T$.
High-symmetry points in the first Brillouin zone are the same as the points for the TBM.
For a later use, we define a spring constant as $K$, and the displacements of the mass points from the equillibrium point as $\phi^{\mu}_{\vec{k},p}$.
Here, $p=1,2,3$ denotes a sublattice and $\mu=x,y$ denotes directions in the two-dimensional space.
Additionally, we define the ratio of a natural length of a spring to the length of the spring in equillibrium as $\eta$
which denotes the tension added to the springs and contributes the interaction between longitudinal and transverse waves.
We note that $\eta$ is the independent parameter in our mechanical model since our system is in equilibrium for any $K$.
In this paper, we focus on the case of the tension for $0 \leq \eta \leq 1$ i.e., the spring is extended by the tension.

With the basis $\vec{\phi} = \left(\phi^{x}_{\vec{k},1},\phi^{y}_{\vec{k},1},\phi^{x}_{\vec{k},2},\phi^{y}_{\vec{k},2},\phi^{x}_{\vec{k},3},\phi^{y}_{\vec{k},3}\right)^T$,
the Newtonian equation of motion describing the vibrations of mass points under the periodic boundary condition is written as
\begin{equation}
 \label{eq:EOM}
 \frac{d^{2}}{dt^{2}}\vec{\phi} = -D (\vec{k}) \vec{\phi},
\end{equation}
where $D(\vec{k})$ is the positive semi-definite matrix called a momentum-space dynamical matrix whose dimension is six since there are three sublattices and two spatial coordinates.
This matrix can be divided into two terms, $D(\vec{k}) = D^{s}(\vec{k}) + D^{h}$.
$D^{s}(\vec{k})$ and $D^h$ are positive semi-definite and positive definite matrices, respectively.
$D^{s}(\vec{k})$ denotes the coupling described by spring constants and tensions of the springs.
However, $D^{s}(\vec{k})$ has the non-uniform diagonal elements.
Therefore, this system does not preserve chiral symmetry because $D^{s}(\vec{k})$ does not anticommute with a chiral operator.

Thus, in order to maintain chiral symmetry, we introduce the on-site potential $D^{h}$ arising from dents on the floor~\cite{PhysRevB.100.054109}.
In this paper, we assume that the shape of the dents is a paraboloid. From the constraint force by the dents, the potential energy due to the gravity is coverted to the vibrational energy. Therefore, mass points oscillate along the surfaces of the dents because of the gravity, which gives rise to the on-site potential [see Fig.~\ref{fig:SMM} (b)].
This on-site potential only affects the diagonal elements of $D(\vec{k})$, and makes diagonal elements of $D(\vec{k})$ constant.
Moreover, this on-site potential has high controllability since the potential depends on the shape of the dents.

Specifically, the explicit form of $D^s(\vec{k})$ and $D^h$ in this system is written as
\begin{equation}
 D^s (\vec{k})=
\begin{pmatrix}
D^s_{11} & D^{s}_{12} & D^s_{13} \\
\left(D^s_{12}\right)^{\dagger} & D^s_{22} & \hat{0}_{2} \\
\left(D^s_{13}\right)^{\dagger} & \hat{0}_{2} & D^s_{33}
\end{pmatrix},
\end{equation}
\begin{equation}
 D^h (\vec{k})=\mathrm{diag}\left(2K \eta,2K\eta,2K,2K(1+\eta ),2K(1+\eta),2K \right),
\end{equation}
with
\begin{subequations}
 \begin{equation}
  D^s_{11}= \mathrm{diag}\left(4K - 2K \eta , 4K -2K \eta \right),
 \end{equation}
 \begin{equation}
  D^s_{22}= \mathrm{diag}\left(2K , 2K(1-\eta) \right),
 \end{equation}
 \begin{equation}
  D^s_{33}= \mathrm{diag}\left(2K(1-\eta) , 2K \right),
 \end{equation}
 \begin{equation}
  D^s_{12}= \mathrm{diag}\left(-K(1+e^{-ik_x}),-K(1+e^{-ik_x})(1-\eta)\right),
 \end{equation}
 \begin{equation}
  D^s_{13}= \mathrm{diag}\left(-K(1+e^{-ik_y})(1-\eta),-K(1+e^{-ik_y})\right).
 \end{equation}
\end{subequations}
Here, $\hat{0}_{n}$ denotes the $n \times n$ zero matrix.
For a later use, we choose $D_{h}$ so that the diagonal elements of $D(\vec{k})$ become $4K$.
We note that there does not exist the coupling between the displacements along $x$ and $y$ direction
since the springs are aligned horizontality along the $x$ and $y$ direction.

Assuming that the mass points oscillate with a frequency $\omega$, one can write $\phi^{\mu}_{\vec{k},p}= e^{-i\omega t} \xi^{\mu}_{\vec{k},p}$.
Substituting this into Eq.~(\ref{eq:EOM}), we obtain
\begin{equation}
\label{eq:eigenvalue equation of SMM}
 -\omega^{2} \vec{\xi}+D(\vec{k}) \vec{\xi}= 0,
\end{equation}
where $\vec{\xi} = \left(\xi^{x}_{\vec{k},1},\xi^{y}_{\vec{k},1},\xi^{x}_{\vec{k},2},\xi^{y}_{\vec{k},2},\xi^{x}_{\vec{k},3},\xi^{y}_{\vec{k},3}\right)^T$
is the eigenvector of $D(\vec{k})$.
As a result, the problem is reduced to the eigenvalue equation analogous to the Schr\"{o}denger equation.
By solving Eq.~(\ref{eq:eigenvalue equation of SMM}), we obtain the dispertion relation.
Figure~\ref{fig:SMM} (c) shows the band structure for $K=1$, $\eta=0.7$.
We note that the momentum space dynamical matrix corresponds the Hamiltonian of the TBM since the SMM is reduced to two copies of the TBM for $\eta=0$, i.e., the tension is infinitely strong.

Let us discuss the symmetry for this mechanical system.
We define a chiral operator $\Upsilon_{S}$ as
\begin{subequations}
 \begin{equation}
\label{eq:Chiral operator S}
  \Upsilon_{S}=\Upsilon_{T} \otimes \hat{1}_{2}  =\mathrm{diag}\left(1,1,-1,-1,-1,-1\right).
 \end{equation}
Additionally, we define a two-fold rotational operator $R_{S}$ around the point \text{\sffamily X} shown in Fig.~\ref{fig:SMM} (a) in this mechanical system as
\begin{equation}
\label{eq:Spatial operator S}
\begin{split}
 R_{S}(\vec{k})&= R_{T}(\vec{k}) \otimes (-\hat{1}_{2}), \\
 &  =\mathrm{diag}\left(-1,-1,-e^{-i k_x},-e^{-i k_x},-e^{-i k_y},-e^{-i k_y}\right).
\end{split}
\end{equation}
\end{subequations}
These operators correspond to the ones for the TBM, respectively.
However, dimensions of these matrix are doubled since the operations are applied not only to the sublattice degrees of freedoms, but also to directions of the displacements.
Two-fold rotational operator is written as $R_{S}(\vec{k})D(\vec{k})R^{-1}_{S}(\vec{k})=D(-\vec{k})$.

The diagonal elements of the momentum-space dynamical matrix just shift the band structure, $\omega^2(\vec{k})$.
Therefore, this system preserves ``chiral symmetry'' since $D(\vec{k})$ subtracted the diagonal elements anticommute with $\Upsilon_{S}$,
$ \{D(\vec{k})-4K \hat{1}_{6},\Upsilon_{S}\}=0$.
Topological zero modes for a SMM corresponding to ones for a TBM can be observed independently of the constant contribution except the shift of these modes. To emphasize this shift, in the rest of this paper, topological zero modes for a SMM are denoted by topological modes.
In addition, this system preserves two-fold rotational symmetry at the high-symmetry points $\Gamma$, $M$, and $X$; $ [D(\vec{k}),R_{S}(\vec{k})]=0$.

From Eq.~(\ref{eq:Chiral operator S}) and Eq.~(\ref{eq:Spatial operator S}), one can confirm that chirality of the lattices is not changed by the operator $R_{S}(\vec{k})$
since $\left[\Upsilon_{S},R_{S}(\vec{k})\right]=0$.
$\Upsilon_{S}$ can be block-diagonalized by the eigenvectors of $R_{S}(\vec{k})$.
Then, the number of the topological modes protected by two-fold rotational symmetry for this system is
\begin{equation}
 \nu=
\begin{cases}
 2 & \mathrm{at} \quad \Gamma \\
 6 & \mathrm{at} \quad M \\
 2 & \mathrm{at} \quad X
\end{cases},
\end{equation}
at least.
The above results can be obtained by noticing $\Upsilon_{S}=\Upsilon_{T} \otimes \hat{1}_{2}$.
We note that the number of topological modes for the SMM is doubled compared to that for the TBM, which is unique properties of the Lieb lattice structure. 
These topological modes are observed as two flat bands at $\Gamma$ and $X$ points, and two flat bands and two Diarc cones at $M$ point in the band structure [see Fig.~\ref{fig:SMM} (c)].
We note that these topological modes are stable for any $\eta$ since $\Upsilon_{S}$ and $R_{S}(\vec{k})$ do not depend on $\eta$.
Therefore, these modes for the SMM are stable even when the longitudinal and the transverse waves are coupled, which demonstrates the robustness of topological modes against the perturbation preserving chiral symmetry.

Additionally, the extra topological modes protected by rotational symmetry are unique topological phenomena for the mechanical system.
These additional topological modes originate from the internal degrees of freedoms (i.e., dispalcements along the $x$- and $y$-direction) which may also shift the eigenspace of $R$ due to the angular momentum.

We finish this part with a remark on generality of our approach. Our approach to realize the topological modes protected by chiral symmetry and rotation symmetry can be applied to generic two-dimensional SMMs. 
Indeed, we can realize the topological modes protected by chiral symmetry and three-fold rotational symmetry for the SMM with a honeycomb structure.

{\bf Summary:}~
We have proposed how to realize the topological modes protected by chiral and rotation symmetry for two-dimensional systems. 
Specifically, by taking advantage of the high controllability of SMMs, we have realized topological modes protected by chiral symmetry and two-fold rotation symmetry. We note that in our approach, introducing the on-site potential arising from the dents on the floor plays an essential role in preserving the chiral symmetry.
In addition, compared to the results of the TBM, the number of topological modes increases for the SMM which is due to the presence of the extra degrees of freedoms.
Our approach by the SMM and the dents on the floor can be applied to a general two-dimensional TBM.

{\bf Acknowledgment:}~
This work is partly supported by JSPS KAKENHI Grants Nos.~JP17H06138, JP20H04627, and JP20K14371.
We thank Senri Suzuki for discussion in an early stage of the work. 


\end{document}